# Surface waves on metal-dielectric boundaries on the frequency of $\omega_{pl}$

V.S.Zuev and G.Ya.Zueva

The P.N.Lebedev Physical Institute of RAS
53 Leninsky pr., 119991 Moscow, Russia
vizuev@sci.lebedev.ru

Abstract
Surface optical plasmons on metal-dielectric boundaries of various shapes are studied. The study features by the exploration of plasmons of the frequency that is larger than $\omega_{pl}/\sqrt{2}$ and approximately equal to $\omega_{pl}$. These plasmons exist on thin films, thin cylinders, and on spheres. Such a plasmon does not exist on a single surface. For applications the use of plasmons of the $\sim \omega_{pl}$ frequency means the promotion of nanophotonics devices into a short wavelength range. For Ag this means the promotion to the wavelength of 140 *nm*.

## *Поверхностные волны на границах раздела металл-диэлектрик на частоте $\omega_{pl}$*

В.С.Зуев, Г.Я.Зуева

Физический институт им. П.Н.Лебедева РАН
119991 Ленинский пр-т, 53, Москва, Россия
vizuev@sci.lebedev.ru

### 1. Плоский слой

Геометрия задачи изображена на рис.1. Пока не будем уточнять, что есть металл, а что – диэлектрик. Будем считать, что $\varepsilon_1 = \varepsilon_3$.

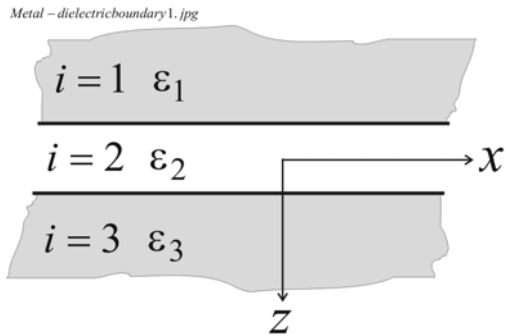

Рис.1.

Речь идет о собственных волнах неоднородного пространства, изображенного на рис.1. Ищем решение в виде поверхностной *TM* - волны. Конечно, это не единственное возможное решение, но именно оно интересует нас. Поле монохроматическое, зависимость от времени $e^{-i\omega t}$. У *TM* - волны магнитное поле имеет только поперечную компоненту $\vec{H} = \vec{a}_y H_y(x,z,t)$, электрическое поле имеет две компоненты: $\vec{E} = \vec{a}_x E_x + \vec{a}_z E_z$. Ищем волну в виде

$$H_y^{(j)}(x,z,t) = H_y^{(j)} e^{-\kappa_z^{(j)} z} e^{ik_x x} e^{-i\omega t}. \tag{1}$$

Плоские волны во всех 3-х средах имеют вид $H_y^{(j)}(x,z,t) = H_y^{(j)} e^{ik_z^{(j)}z} e^{ik_x x} e^{-i\omega t}$. Для простоты зависимость от $y$ не вводим. Все магнитные проницаемости $\mu_i = 1$. Для всех сред имеет место соотношение $k_x^2 + k_z^2 = (\omega/c)^2 \varepsilon$. Это равенство есть дисперсионное уравнение для плоских волн в однородной (в том числе – в некоторой части пространства) среде. У поверхностных волн $k_z^2 = -\kappa_z^2 < 0$. При извлечении корня возникают два знака. Выбор знака определяется выбором направления затухания волны вдоль оси $z$.

В задаче имеется четыре волны вида (1), причем все они имеют одно и то же $k_x$. Это условие возникает вследствие граничных условий равенства тангенциальных компонент полей. В среде $i = 1$ имеется одна волна с индексом $j = 1$, убывающая в направлении $z \to -\infty$. У этой волны $k_z^{(j=1)} = -i\sqrt{k_x^2 - (\omega/c)^2 \varepsilon_1} = -i\kappa_{1z}$. В среде $i = 2$ имеются две волны $j = 2$ и $j = r$, $k_z^{(j=2)} = i\sqrt{k_x^2 - (\omega/c)^2 \varepsilon_2} = i\kappa_{2z}$ и $k_z^{(j=r)} = -i\sqrt{k_x^2 - (\omega/c)^2 \varepsilon_2} = -i\kappa_{2z}$. Волна $j = 2$ затухает в направлении $z \to \infty$, волна $j = r$ затухает в направлении $z \to -\infty$,. В среде $i = 3$ имеется одна волна, $k_z^{(j=3)} = i\sqrt{k_x^2 - (\omega/c)^2 \varepsilon_3} = i\kappa_{3z}$. Она затухает в направлении $z \to \infty$,. Выбрав $\varepsilon_3 = \varepsilon_1$ получаем $\kappa_{3z} = \kappa_{1z}$. Выпишем волны (зависимость от времени опустим):

$$H_y^{(1)}(x,z) = H_{1y} e^{\kappa_{1z}z} e^{ik_x x}, \quad H_y^{(2)}(x,z) = H_{2y} e^{-\kappa_{2z}z} e^{ik_x x}, \quad (2)$$
$$H_y^{(r)}(x,z) = H_{ry} e^{\kappa_{2z}z} e^{ik_x x}, \quad H_y^{(3)}(x,z) = H_{3y} e^{-\kappa_{1z}z} e^{ik_x x}.$$

Электрическое поле получим с помощью уравнения $\nabla \times \vec{H} + i\dfrac{\omega}{c}\varepsilon \vec{E} = 0$, входящего в систему уравнений Максвелла для пространства без свободных зарядов и токов. Нам потребуются только $E_x$-компоненты. В $TM$ - волне они вычисляются по формуле $E_{jx} = -i\dfrac{c}{\omega \varepsilon_i}\dfrac{d}{dz} H_{jy}$ и имеют вид:

$$E_{1x} = -i\frac{c\kappa_{1z}}{\omega \varepsilon_1} H_{1y} e^{\kappa_{1z}z} e^{ik_x x}, \quad E_{2x} = i\frac{c\kappa_{2z}}{\omega \varepsilon_2} H_{2y} e^{-\kappa_{2z}z} e^{ik_x x}, \quad (3)$$
$$E_{rx} = -i\frac{c\kappa_{2z}}{\omega \varepsilon_2} H_{ry} e^{\kappa_{2z}z} e^{ik_x x}, \quad E_{3x} = i\frac{c\kappa_{1z}}{\omega \varepsilon_1} H_{3y} e^{-\kappa_{1z}z} e^{ik_x x}.$$

Будем считать, что границы среды $i = 2$ расположены при $z = 0$ и $z = d$. Составляем граничные условия равенства тангенциальных компонент полей на этих границах.

$$H_{1y} - H_{2y} - H_{ry} + 0 \cdot H_{3y} = 0$$
$$H_{1y} + \frac{\kappa_{2z}\varepsilon_1}{\kappa_{1z}\varepsilon_2} H_{2y} - \frac{\kappa_{2z}\varepsilon_1}{\kappa_{1z}\varepsilon_2} H_{ry} + 0 \cdot H_{3y} = 0 \quad (4)$$
$$0 \cdot H_{1y} + e^{-\kappa_{2z}d} H_{2y} + e^{\kappa_{2z}d} H_{ry} - e^{-\kappa_{1z}d} H_{3y} = 0$$
$$0 \cdot H_{1y} + \frac{\kappa_{2z}\varepsilon_1}{\kappa_{1z}\varepsilon_2} e^{-\kappa_{2z}d} H_{2y} - \frac{\kappa_{2z}\varepsilon_1}{\kappa_{1z}\varepsilon_2} e^{\kappa_{2z}d} H_{ry} - e^{-\kappa_{1z}d} H_{3y} = 0$$

Вводим обозначение $K = -\dfrac{\kappa_{2z}\varepsilon_1}{\kappa_{1z}\varepsilon_2}$. Составляем детерминант этой системы и приравниваем его нулю. Возникает характеристическое уравнение.

$$(K-1)^2 - (K+1)^2 e^{-2\kappa_{2z}d} = 0 \tag{5}$$

Квадратное относительно $K$ уравнение имеет следующие корни:

$$-\frac{\kappa_{2z}\varepsilon_1}{\kappa_{1z}\varepsilon_2} = \frac{1+e^{-\kappa_{2z}d}}{1-e^{-\kappa_{2z}d}}, \ \frac{1-e^{-\kappa_{2z}d}}{1+e^{-\kappa_{2z}d}} \tag{6}$$

Таким образом характеристическое уравнение представляет собой произведение двух сомножителей, каждый из которых может быть равен нулю:

$$\left(\frac{\kappa_{2z}\varepsilon_1}{\kappa_{1z}\varepsilon_2} + \frac{1+e^{-\kappa_{2z}d}}{1-e^{-\kappa_{2z}d}}\right)\left(\frac{\kappa_{2z}\varepsilon_1}{\kappa_{1z}\varepsilon_2} + \frac{1-e^{-\kappa_{2z}d}}{1+e^{-\kappa_{2z}d}}\right) = 0 \tag{7}$$

Все входящие в (7) величины кроме $\varepsilon_1$ и $\varepsilon_2$ положительны. Решение существует лишь в том случае, если одно из $\varepsilon$ отрицательно. Будем поочередно полагать $\varepsilon_1 < 0$ и $\varepsilon_2 < 0$.

Результат расчета показан на рис.2. Обозначения на рис.2 имеют следующий смысл:

$$srs(x,k0,\varepsilon1,\varepsilon2) = -\frac{\varepsilon_2}{\varepsilon_1}\frac{1+e^{-d\sqrt{k_x^2-k_0^2\varepsilon_2}}}{1-e^{-d\sqrt{k_x^2-k_0^2\varepsilon_2}}}, \ ars(x,k0,\varepsilon1,\varepsilon2) = -\frac{\varepsilon_2}{\varepsilon_1}\frac{1-e^{-d\sqrt{k_x^2-k_0^2\varepsilon_2}}}{1+e^{-d\sqrt{k_x^2-k_0^2\varepsilon_2}}},$$

$$ls(x,k0,\varepsilon_1,\varepsilon_2) = \frac{\sqrt{k_x^2-k_0^2\varepsilon_2}}{\sqrt{k_x^2-k_0^2\varepsilon_1}}, \ x \text{ соответствует } k_x. \text{ Толщина слоя } d \text{ равна } 2 \ nm, \text{ волновое число}$$

$k_0 = \dfrac{2\pi}{\lambda_0}$ принимает два значения $k_{l0}$ и $k_{h0}$, соответствующие длинам волн $245 \ nm$ и $172 \ nm$. Такой выбор значений $\varepsilon$ для металла и длин волн соответствует серебру. Фрагменты $a$ и $b$ относятся к слою металла в диэлектрике, фрагменты $c$ и $d$ - к слою диэлектрика в металле.

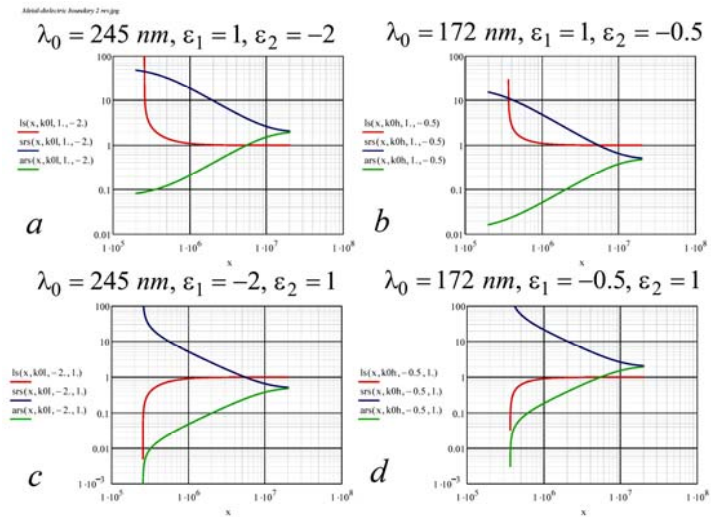

Рис.2.

Следует считать, что на фрагментах $c$ и $d$ красные линии пересекаются с зелеными линиями и при малых значениях $k_x$ (на рисунке - $x$).

Синие и зеленые кривые на рис.2 соответствуют двум ветвям характеристического уравнения (7), которое состоит из двух сомножителей. Рис. 2 демонстрирует два удивительных свойства. Во-1-ых, вопреки тому, что имеет место для одиночной границы раздела металл-диэлектрик, где поверхностные волны существуют лишь при условии $\varepsilon_{metal} \leq -1$, на тонких пленках поверхностные волны существуют и при $-1 < \varepsilon_{metal} < 0$. Во-2-ых, при $-1 < \varepsilon_{metal} < 0$ два корня, т. е. два плазмона возникают для одной и той же ветви характеристического уравнения. В случае, изображенном на фрагменте $a$, различные ветви характеристического уравнения соответствовали симметричному и антисимметричному плазмону. Высказанные в данном абзаце соображения относятся к случаю $\varepsilon_{dielectric} = 1$.

## 2. Сфера

Начнем с волнового уравнения.

$$\nabla\nabla\cdot\vec{C} - \nabla\times\nabla\times\vec{C} + k^2\vec{C} = 0, \qquad (1)$$
$$k^2 = \varepsilon\mu\omega^2.$$

Уравнение (1) имеет решения вида

$$\vec{L} = \nabla\psi, \quad \vec{M} = \nabla\times\vec{a}\psi, \quad \vec{N} = \frac{1}{k}\nabla\times\vec{M}. \qquad (2)$$

Здесь $\psi$ - решение уравнения

$$\nabla^2\psi + k^2\psi = 0, \qquad (3)$$

$\vec{a}$ - постоянный вектор единичной длины, разный в разных координатных системах.

В сферических координатах $r, \theta, \varphi$ решение уравнения (3) имеет вид $\psi_{\substack{e\\o}mn} = f_{\substack{e\\o}mn} e^{-i\omega t}$, где

$$f_{\substack{e\\o}mn} = {\substack{\cos\\\sin}} m\varphi \, P_n^m(\cos\theta) z_n(kR). \qquad (4)$$

Следует допускать, что $k$ может быть мнимой величиной.

Характеристическое уравнение для собственных волн сферы имеет вид:

$$\frac{[N\rho j_n(N\rho)]'}{N^2 j_n(N\rho)} = \frac{\mu_2}{\mu_1}\frac{[\rho y_n(\rho)]'}{y_n(\rho)}. \qquad (5)$$

$k^2 = \varepsilon\mu\omega^2$, $N = \frac{k_1}{k_2} = \sqrt{\frac{\varepsilon_1}{\varepsilon_2}}$, $\rho = k_2 a = a\omega\sqrt{\varepsilon_2\mu_0}$, $N\rho = a\omega\sqrt{\varepsilon_2\mu_0}\sqrt{\frac{\varepsilon_1}{\varepsilon_2}} = a\omega\sqrt{\varepsilon_1\mu_0}$.

Система единиц – как у Стрэттона, $M.K.S.$ или $Giorgi\ system$, в этой системе единиц скорость света $c = 1/\sqrt{\varepsilon_0\mu_0}$, и поэтому $k^2$ имеет такой странный вид. $\varepsilon_1$ относится к среде внутри сферы, $\varepsilon_2$ - к среде вне сферы, $\mu_1 = \mu_2 = \mu_0$ в единицах Стрэттона. Обозначим $a_0 = a\omega\sqrt{\varepsilon_0\mu_0}$, $\varepsilon_1/\varepsilon_0 = \varepsilon_{01}$, $\varepsilon_2/\varepsilon_0 = \varepsilon_{02}$. После вычисления производных формула (5) принимает вид:

$$\frac{N\rho\, j_{n-1}(N\rho) - n j_n(N\rho)}{j_n(N\rho)} = N^2 \frac{\rho\, y_{n-1}(\rho) - n y_n(\rho)}{y_n(\rho)}. \qquad (6)$$

Величины $\varepsilon_{01}$ и $\varepsilon_{02}$ теперь стали безразмерными, и $a_0 = a(\omega/c)$, как в системе CGSE. Функция (6) – либо функция $a_0$, либо функция $\varepsilon_{01}$, либо $\varepsilon_{02}$.

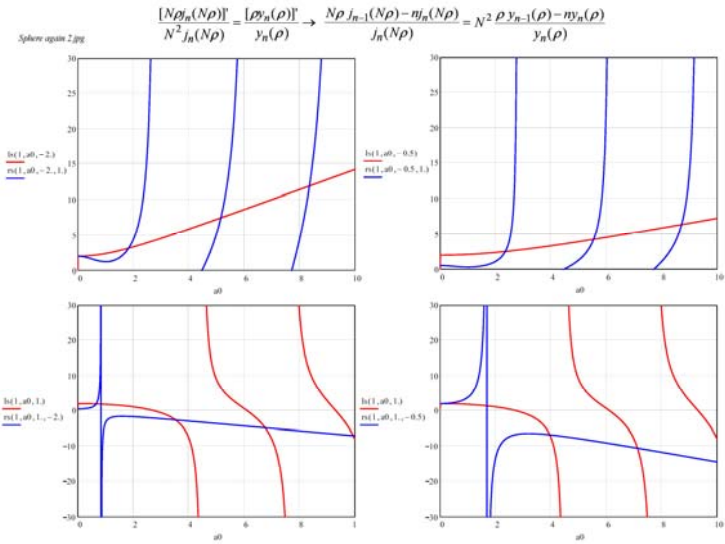

$$\frac{[N\rho j_n(N\rho)]'}{N^2 j_n(N\rho)} = \frac{[\rho y'_n(\rho)]'}{y_n(\rho)} \rightarrow \frac{N\rho j_{n-1}(N\rho) - nj_n(N\rho)}{j_n(N\rho)} = N^2 \frac{\rho y_{n-1}(\rho) - ny_n(\rho)}{y_n(\rho)}$$

Нам известно лишь одно указание на существование поверхностных волн на границе металл-диэлектрик при $\varepsilon_{diel} = 1$, $-1 > \varepsilon_{met} > 0$. Это указание содержится в работе /1/. Проделанный выше анализ выявил условия существования поверхностных волн при $\varepsilon_{diel} = 1$, $-1 > \varepsilon_{met} > 0$. С практической точки зрения это означает, что устройства нанофотоники с поверхностными волнами могут работать при длине волны в $\sqrt{2}$ раз более короткой, чем можно было бы думать, если пользоваться известной формулой для волнового числа поверхностного плазмона на одиночной границе раздела $k_{pl} = k_0 \sqrt{\dfrac{\varepsilon_1 \varepsilon_2}{\varepsilon_1 + \varepsilon_2}}$ и плазменным законом дисперсии для металла $\varepsilon_{met} = 1 - (\omega_{pl}/\omega)^2$. Для серебра это означает продвижение вплоть до длины волны $140\ nm$.

1. D.V.Guzatov and V.V.Klimov. arXiv:physics/0703251 28 Mar 2007